\documentclass[runningheads]{llncs}
\usepackage{graphicx}
\usepackage{amsmath}
\usepackage{amssymb}
\usepackage{bm}

\begin{document}
\title{GANBERT: Generative Adversarial Networks with Bidirectional Encoder Representations from Transformers for MRI to PET synthesis}
\titlerunning{GANBERT for MRI to PET synthesis}
%
\author{Hoo-Chang Shin\inst{1}, Alvin Ihsani\inst{1}, Swetha Mandava\inst{1}, Sharath Turuvekere Sreenivas\inst{1}, Christopher Forster\inst{1}, 
Jiook Cha\inst{2}\\ \and Alzheimer's Disease Neuroimaging Initiative} 


\authorrunning{Shin et al.}
\institute{NVIDIA Corporation;
\email{hshin@nvidia.com}\and
Department of Psychology, Center for REAL Intelligence, AI Institute, Seoul National University;
\email{connectome@snu.ac.kr}}
%
\maketitle              
\begin{abstract}

Synthesizing medical images, such as PET, is a challenging task due to the fact that the intensity range is much wider and denser than those in photographs and digital renderings and are often heavily biased toward zero.
Above all, intensity values in PET have absolute significance, and are used to compute parameters that are reproducible across the population.
Yet, usually much manual adjustment has to be made in pre-/post- processing when synthesizing PET images, because its intensity ranges can vary a lot, e.g., between \texttt{-100}\,-\,\texttt{1000} in floating point values.

To overcome these challenges, we adopt the Bidirectional Encoder Representations from Transformers (BERT) algorithm that has had great success in natural language processing (NLP), where wide-range floating point intensity values are represented as integers ranging between \texttt{0}\,-\,\texttt{$10^4$} that resemble a dictionary of natural language vocabularies.
BERT is then trained to predict a proportion of masked values images, where its ``next sentence prediction (NSP)'' acts as GAN discriminator.

Our proposed approach, is able to generate PET images from MRI images in wide intensity range, with no manual adjustments in pre-/post- processing.
It is a method that can scale and ready to deploy.

\end{abstract}

\section{Introduction}

Positron emission tomography (PET) is a nuclear medicine functional imaging technique that is used to diagnose a wide range of diseases including Alzheimer Disease (AD) in early and late stages~\cite{Johnson16}.
In PET imaging, a radioactive tracer is injected in the body, followed by a scan that measures the high-energy photons emitted from the tracer as it travels through the body.
The projections (or photon counts) acquired during the scan are reconstructed into (single- or multi-frame) 3D volumes that allow for the visualization and analysis of the tracer and its binding properties in tissues.

There are three types of widely used PET imaging techniques: \texttt{AV45} that measures amlyloid uptake; \texttt{AV1451} that measures tau protein aggregates; and \texttt{FDG} that best visualizes the patterns of glucose metabolism.
While PET imaging provides great insight into the functional processes for which the radiotracer is developed in the brain or body, it provides no anatomical information and therefore MR (magnetic resonance) or CT (computed tomography) images must be acquired.
T1-weighted MRI imaging technique best highlights the body anatomy, which is mostly the type of MRI when performed together with PET.

\begin{figure}[t]
\centering
\includegraphics[width=.8\linewidth]{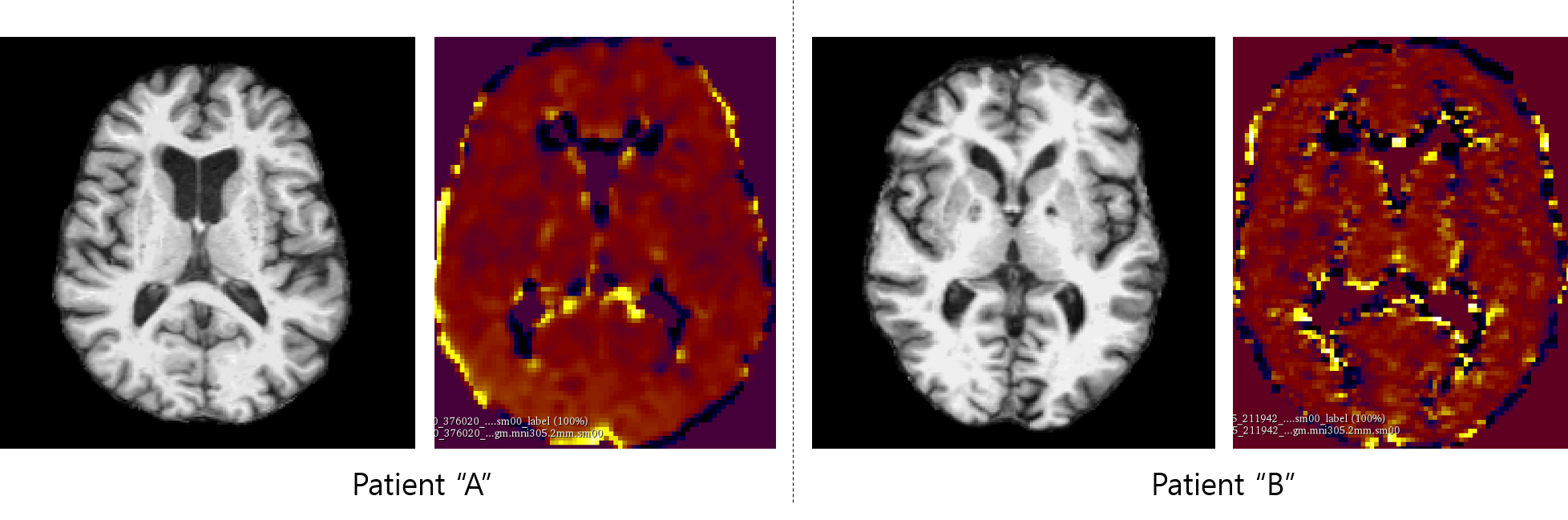}
\caption{Examples of T1-weighted MRI and PET images of same patient. T1 images are in $256\times 256\times 256$ size and PET are $N\times 93\times 76\times 76$ size where N is the number of time-points PET is taken. It is challenging to register MRI and PET images.}
\label{fig:t1_mri_example}
\end{figure}

\begin{figure}[t]
\centering
\includegraphics[width=1\linewidth]{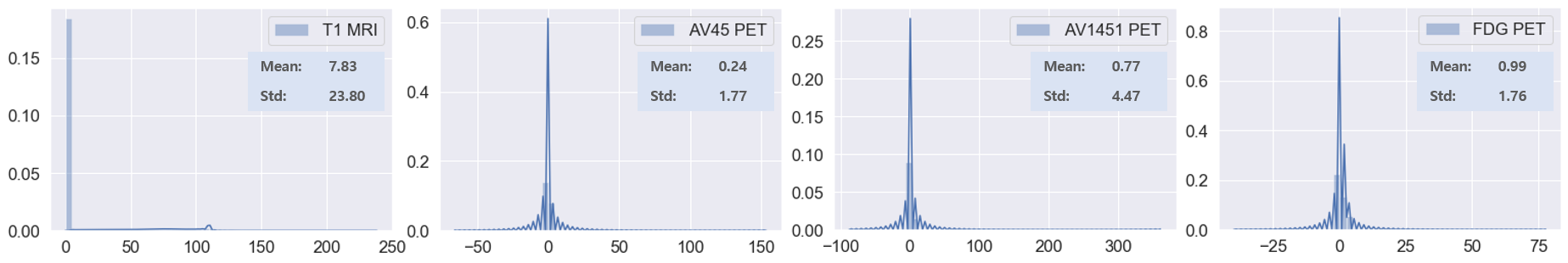}
\caption{Distribution of the image intensity values in T1-Weighted MRI, AV45, AV1451, and FDG PET images on five random samples. Values are highly centered around zero, with wide intensity range.}
\label{fig:data_distribution}
\end{figure}

In this paper, we explore the question of synthesizing PET images from existing MR images.
This is a challenging task by nature, because we are expecting to see radioactivity over time from MRI that does not have any.
Unlike semantic segmentation where the ``paired''~\cite{isola2017image,zhu2017unpaired} images (e.g., images and sketches, label maps) mostly have perfect matching, MRI and PET images look quite different even though imaging the same object.
An example of MRI and corresponding PET is shown in Figure~\ref{fig:t1_mri_example}.
It is also a difficult task from a deep learning perspective, since PET imaging can result in images whose intensities range between -100 to 1000 or more, in floating point values, as shown in Figure~\ref{fig:data_distribution}.

\section{Related Works}

We demonstrate synthesizing the PET in Standardized Uptake Value Ratio (SUVR)~\cite{zasadny1993standardized} without subject-specific prior knowledge or manual adjustments.
SUVR is the most commonly used to quantitatively analyze the degree of radiotracer uptake, computed by comparing a reference region with the other regions e.g., using segmentation maps.
PET is usually scanned over time-steps, typically 2 to 4, and in this study we generate PET with 2 time-steps based on MRI input.

\subsection{MRI-PET and Other Cross-Modal Medical Image Synthesis}

PET synthesis from MRI using GAN was demonstrated in \cite{yan2018generation} for AD classification.
A total of 108 amyloid (AV45) PET and the corresponding MRI image pairs with 58 early Mild Cognitive Impairment (EMCI) and 50 stable Mild Cognitive Impairment (MCI) from the publicly available Alzheimer’s Disease Neuroimaging Initiative (ADNI) dataset.
However, it is unclear from the paper which values were used to synthesize PET in what intensity range.

MRI images with different magnetic pulse sequences, namely T1 and FLAIR, were synthesized using unpaired image-to-image translation GAN~\cite{zhang2019harmonic}.
T1 to FLAIR has very good one-to-one mapping (or registration) while MRI to PET does not.
There are many other works that generate computerized tomography (CT) scan from MRI~\cite{lei2019mri,yang2018unpaired} or vice-versa~\cite{rubin2019ct}.
Most of them use the Pix2Pix-GAN~\cite{isola2017image} or Cycle-GAN~\cite{zhu2017unpaired}.
MRI and CT can be well-represented by 0-255 \texttt{uint8} range, and they also can have very good registration.
Applying the Pix2Pix-GAN for example does not give good result for our problem of generating PET generation in its SUVR values from MRI.
The purpose of the Cycle-GAN is mainly to achieve style transfer without having a ``paired'' dataset.

MRI was generated from amyloid (AV45) PET in~\cite{choi2018generation} using 163 and 98 PET/MRI pairs from the ADNI dataset was used for training/testing, respectively.
Pix2Pix-GAN was used to generatge MRI that was normalized to -1 to 1 intensity range.
Similarly to this work, minimally processed PET images of different scanners were used in our work to generate PET images.

\subsection{PET synthesis from Ultra-low-dose PET}

Generating PET from ultra-low-dose amyloid PET using deep learning was demonstrated in \cite{ouyang2019ultra,chen2019ultra}.
Reconstructing PET from ultra-low-dose amyloid PET, they report reporting $\sim$32  peak signal-to-noise ratio (PSNR), $\sim$0.90 structural similarity (SSIM), and $\sim$0.33 root mean square error (RMSE).
Meanwhile, the range of the SUVR values reconstructed is unclear from the papers.



\section{Dataset}

\subsection{Alzheimer Disease Neuroimaging Initiative Database}

We use the publicly available ADNI\footnote{\url{http://adni.loni.usc.edu}} database.
From the ADNI dataset we collected PET scans with three different imaging protocols that have paired T1-weighted MRI images: 2,387 Amyloid PET (AV45), 536 Tau PET (AV1451), and 3,108 fluorodeoxyglucose PET (FDG).

\subsubsection{Restoring Original PET Values with MRI Statistics}

We normalize the MRI image by subtracting with its \texttt{mean} and dividing by \texttt{standard deviation}.
On the other hand, PET images are normalized by the corresponding MRI statistics, subtracted by 1/10 of the corresponding MRI images' \texttt{mean} and divided by 1/10 of MRI image's \texttt{standard deviation}.
We do so, in order to restore PET's original intensity values.
When PET images are normalized with the corresponding MRI input images, then the PET original intensity values can be most closely restored using the MRI statistics.
On the contrary, if normalized by their own PET image statistics then the original value ranges get lost.

\section{Methods}

\begin{figure}
\centering
\includegraphics[width=1\linewidth,trim=6.2cm 2.8cm 5.8cm 2.8cm,clip]{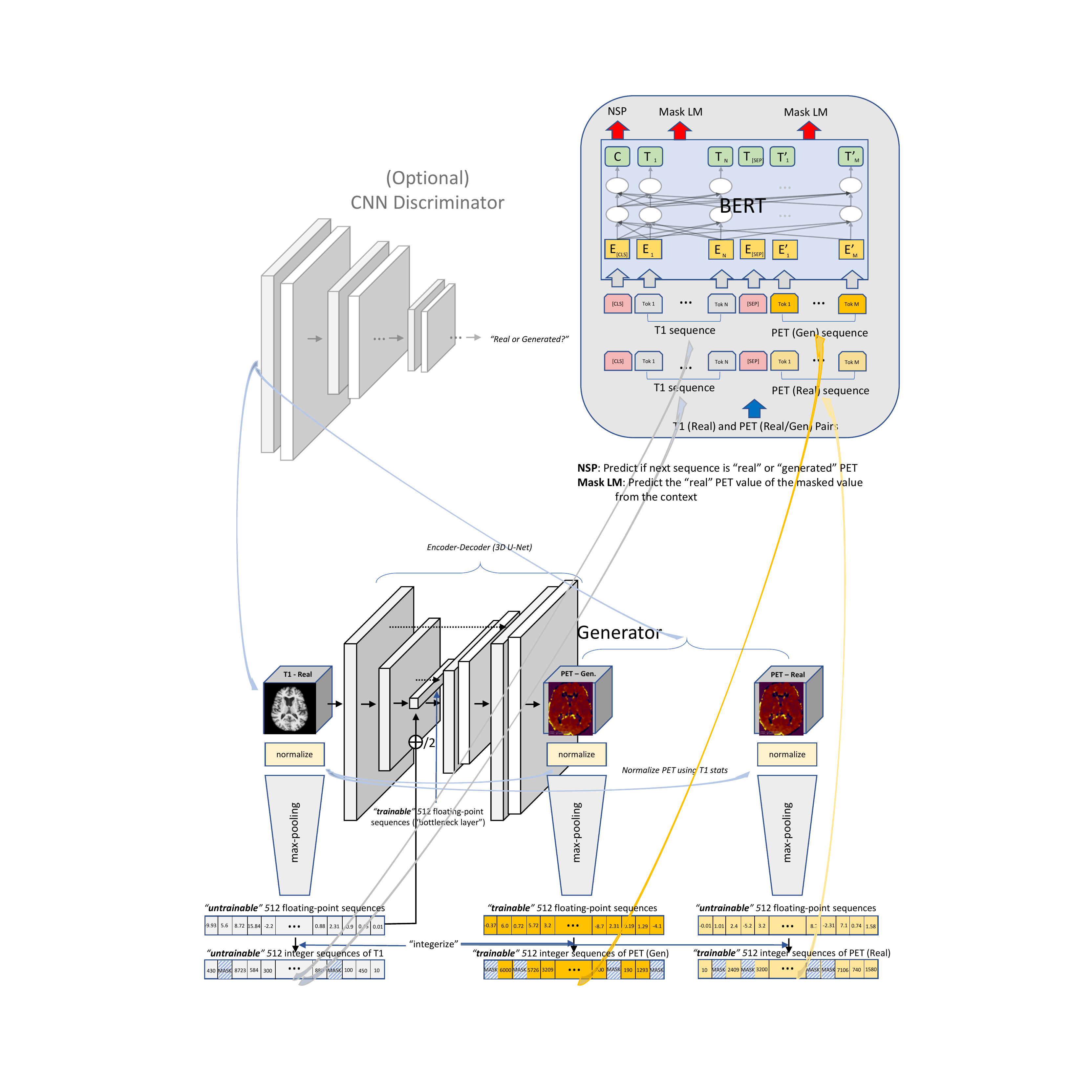}
\caption{Overall architecture and training procedures for GANBERT. U-Net-like architecture gets T1-MRI as input and generates PET image. All (T1/PET(Gen)/PET(Real)) the images are ``summarized'' into 512-length sequences using max-pooling over the image volumes. The sequences are then ``quantized'' by multiplied with $10^3$ and subsequently rounded. The U-Net bottleneck layer is added with the T1-MRI ``summarized'' input sequence and divided by 2. BERT takes the ``integerized'' sequences and is trained as if these were natural language text sequences. `Next sentence prediction' (NSP) task acts as the GAN discriminator, and `masked language model' predicts absolute maximum values in generated PET image sub-regions that are ``masked''.}
\label{fig:ganbert}
\end{figure}

The overall architecture consists of \textit{1} 3D U-Net~\cite{ronneberger2015u}-like generator synthesizing PET from T1-weighted MRI input, and \textit{2} Bidirectional Encoder Representations from Transformers (BERT)~\cite{devlin2018bert} that is trained to predict real vs. generated PET (NSP), and to predict masked values (MLM) from the max-pooled images.
The overall architecture and training procedure are shown in Figure~\ref{fig:ganbert}.

\subsection{U-Net Architecture Generator}

The U-Net like architecture generates PET from T1-MRI input in 3D image space, where the overall loss is sum of the discriminator and L1 losses. 
%
%
The encoder part takes ($256\times 256\times 256$) size T1-MRI image and decoder generates 4D PET images ($2\times 93\times 76\times 76$) that are 3D images over two time-steps.
%
%

We also replace the final layer activation of the generator (decoder) with \texttt{tanhshrink} (Figure~\ref{fig:tanhshrink} in Appendices)  from \texttt{tanh} activation that was used in the Pix2Pix-GAN, to adopt for \texttt{-100}\,-\,\texttt{1000} in floating point value output.
Yet, it is still easy for the generator to generate small values and settle there as local optima.
The ``masked language model'' (MLM) and loss of BERT followed by max-pooling summarization encourages the generator to generate wider range high-intensity values.

\subsection{Bidirectional Encoder Representations from Transformers}

BERT was introduced in~\cite{devlin2018bert}, where it achieved new state-of-the-art on 11 popular NLP tasks.
The main ideas of BERT are \textit{(1)} self-attention through bidirectional transformers, and \textit{(2)} pre-training on large scale data with masked language model (MLM) and next sentence prediction (NSP).

Our hypothesis is that this self-attention over MRI and PET is helpful to synthesize PET from MRI when trained on enough samples, even though the input MRI image itself does not have any radioactive tracer information.
We use the the \textbf{BERT\textsubscript{BASE}} model with 12 layers, 768 hidden size, and 12 attention heads.

\subsubsection{Summarizing Images to ``Text-Like'' Sequences}



The images are ``summarized'' to text-like sequences to be trained with NSP and MLM training objectives of BERT.
Max-pooling is applied so images are summarized with the absolute maximum values in each sub-voxel.
The MRI and PET sequences are then concatenated, separated by \texttt{[SEP]} special token and with additional \texttt{[CLS]} special token at the beginning and end of the entire sequence.
The total length of the sequence is therefore 1027 ($1 + 512 + 1 + 512 + 1$).


The floating-point summarized values are then multiplied with $10^3$ and rounded to integers ranging from 1 to $10^4$, that are like a dictionary with $10^4$ vocabularies.
Negative values are multiplied by -1 and get the remainder dividing by 500, and similarly for the values that are greater than $10^4$, but without flipping its sign and adding 500.
This is similar to the original BERT where the first $\sim$1000 words in the vocabulary dictionaries are \texttt{[UNKNOWN]} special tokens.

\subsubsection{Next Sentence Prediction}

The original BERT paper~\cite{devlin2018bert} trains 50\% of the time with actual next sentence, and 50\% of the time with a random sentence from the training text corpus.
We adopt this idea for our PET synthesis pipeline so it acts as the \textit{discriminator} of GAN, predicting if an image is generated or real.
In the same way the original BERT is trained, real and generated images are selected 50\% of the time each.

\subsubsection{Masked Language Model}

We mask 5\% of the MRI and 25\% of the PET, both generated and real ones.
For generated PET, BERT is tasked to predict the real PET numbers given the 95\% MRI and 75\% of generated sequences.
The proportion of the masked values in the sequences play an important role in the training and the final performance of the model.
If the proportion of the masked sequence is too large, say, over 50\% for example, it becomes too difficult for the discriminator (NSP) to predict real- vs. generated- PET sequences because most of the values are ``masked''.
On the other hand, if the proportion is too small, then predicting the masked values (MLM) will be too difficult.

\subsection{Overall Training Objective}

We train the generator with adversarial loss that is the NSP loss of BERT, $L_1$ loss for image similarity between the generated and the real PET, and MLM loss that lets BERT to predict the correct real PET values from the masked ones.
Each is multiplied with hyper-parameter $\lambda$, such that

\begin{align}
    G^*  = \lambda_{NSP}\arg\min_G\max_D \mathcal{L}_{cGAN}(G,D) + \lambda_{MLM}\mathcal{L}_{MLM}(G) + \lambda_{L1} \mathcal{L}_{L1}(G),\label{full_objective}
\end{align}

\noindent
where $D$ above is BERT in our case that has NSP and MLM objectives, and we set $\lambda_{NSP}=20$, $\lambda_{MLM}=1$, and $\lambda_{L1}=20$.
Similarly to GAN where the generator (G) and the discriminator (D) are trained separately with \texttt{minimax} objective, our generator and BERT are trained separately, each having their own optimizer.


\section{Results}

\setlength{\tabcolsep}{4pt}
\begin{table}
\begin{center}
\caption{Evaluation measures of the generated AV45-PET using different methods: \textit{(1)} \texttt{pix2pix}-GAN with \texttt{tanh} activation; \textit{(2)} using \texttt{tanhshrink} activation to accommodate wider intensity value range; \textit{(3)} GANBERT without additional CNN discriminator (CNN-D); and \textit{(4)} with additional CNN-D. The metrics are measured on PET image generated in -10 to 1000 range floating-point values.}
\label{table:results_av45_pet_measure}
\begin{tabular}{|l||ccc|}
\hline
\noalign{\smallskip}
 & \multicolumn{3}{c|}{AV45-PET} \\
\noalign{\smallskip}
\hline
\noalign{\smallskip}
 Method  & PSNR & SSIM & RSME  \\
\noalign{\smallskip}
\hline\hline
\noalign{\smallskip}
pix2pix (tanh) & 50.41 & 0.0 & 1.85 \\
pix2pix (tanhshrink) & 48.66 & 0.0 & 2.28 \\
GANBERT (no CNN-D) & 57.58 & 0.27 & 0.80 \\
GANBERT (with CNN-D) & 56.53 & 0.31 & 0.91 \\
\hline
\end{tabular}
\end{center}
\end{table}

\setlength{\tabcolsep}{4pt}
\begin{table}
\begin{center}
\caption{Evaluation measures of the generated AV1451-/FDG- PET images.}
\label{table:results_av1451_fdg_pet_measure}
\begin{tabular}{|l||ccc|ccc|}
\hline\noalign{\smallskip}
 & \multicolumn{3}{c|}{AV1451-PET} & \multicolumn{3}{c|}{FDG-PET} \\
\noalign{\smallskip}
\hline
\noalign{\smallskip}
 Method  & PSNR & SSIM & RSME & PSNR & SSIM & RSME \\
\noalign{\smallskip}
\hline\hline
\noalign{\smallskip}
GANBERT (no CNN-D) & 51.05 & 0.25 & 0.82 & 56.53 & 0.11 & 0.82 \\
GANBERT (with CNN-D) &  51.25 & 0.26 & 0.81 & 47.52 & 0.38 & 2.38 \\
\hline
\end{tabular}
\end{center}
\end{table}

\subsection{Quantitative Evaluation}

Similarly to the prior works that reconstructed PET SUVR values from ultra-low-dose PET~\cite{ouyang2019ultra,chen2019ultra}, we evaluate the quality of the generated PET \textit{(1)} using Peak Signal-to-Noise Ratio (PSNR; higher is better), \textit{(2)} Structural Similarity Index (SSIM; higher is better)~\cite{wang2004image}, and \textit{(3)} Root-Mean-Square Error (RMSE; lower is better).
%
The evaluation results of the generated PET images are shown in Table~\ref{table:results_av45_pet_measure} and Table~\ref{table:results_av1451_fdg_pet_measure}.
Using usual 3D \texttt{pix2pix} architecture, it is impossible to reconstruct the original PET SUVR image.
This is also the case when using the \texttt{tanhshrink} activation function in the generator to accommodate the wide intensity range values.
%
It is also noticeable that reasonable images can be generated by training the generator with BERT alone, but also with additional CNN discriminator (CNN-D).
Overall, images with higher SSIM is observed when trained with additional CNN-D though having higher RSME.

\begin{figure}[t]
\centering
\includegraphics[width=.7\linewidth]{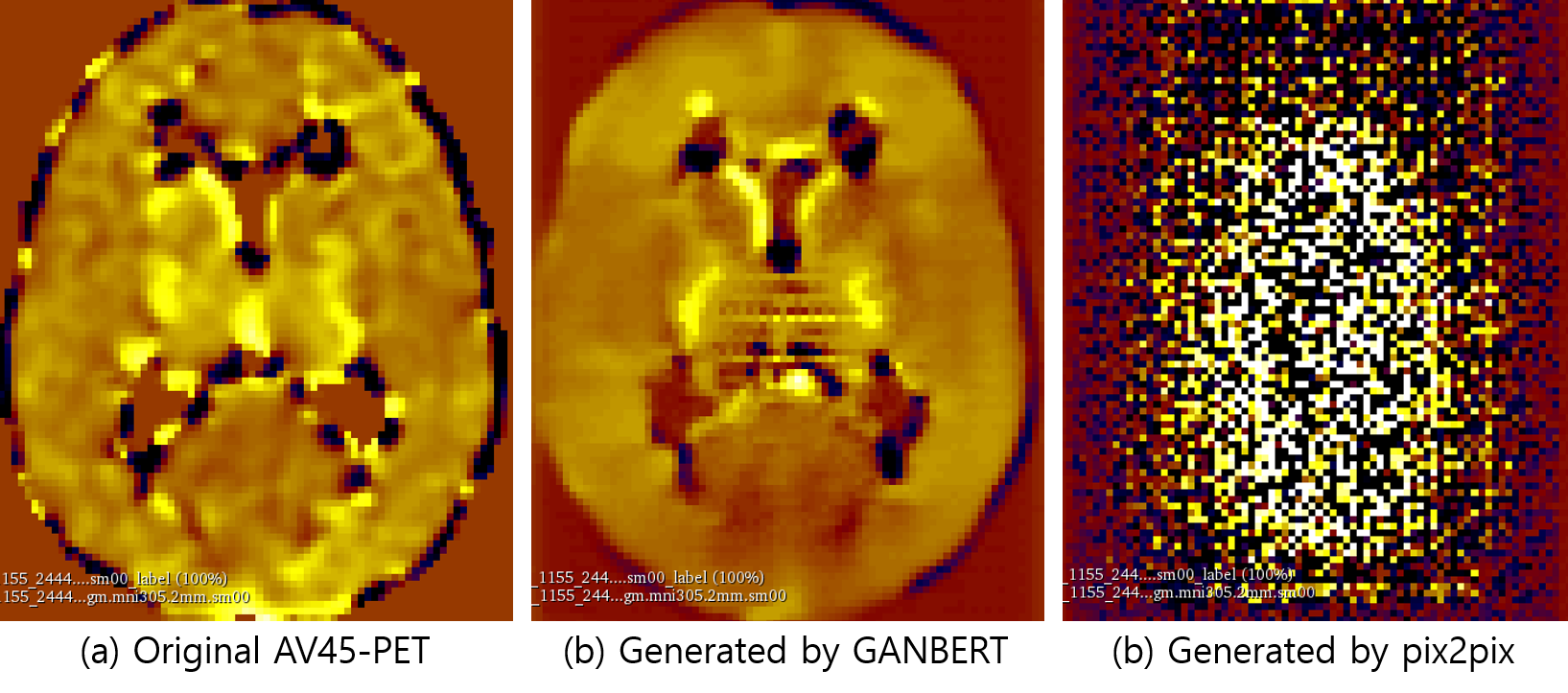}
\caption{AV45-PET (a) original; synthesized by (b) GANBERT; and (c) 3D-\texttt{pix2pix}.}
\label{fig:av45_pet_gen}
\end{figure}

\begin{figure}[t]
\centering
\includegraphics[width=.8\linewidth]{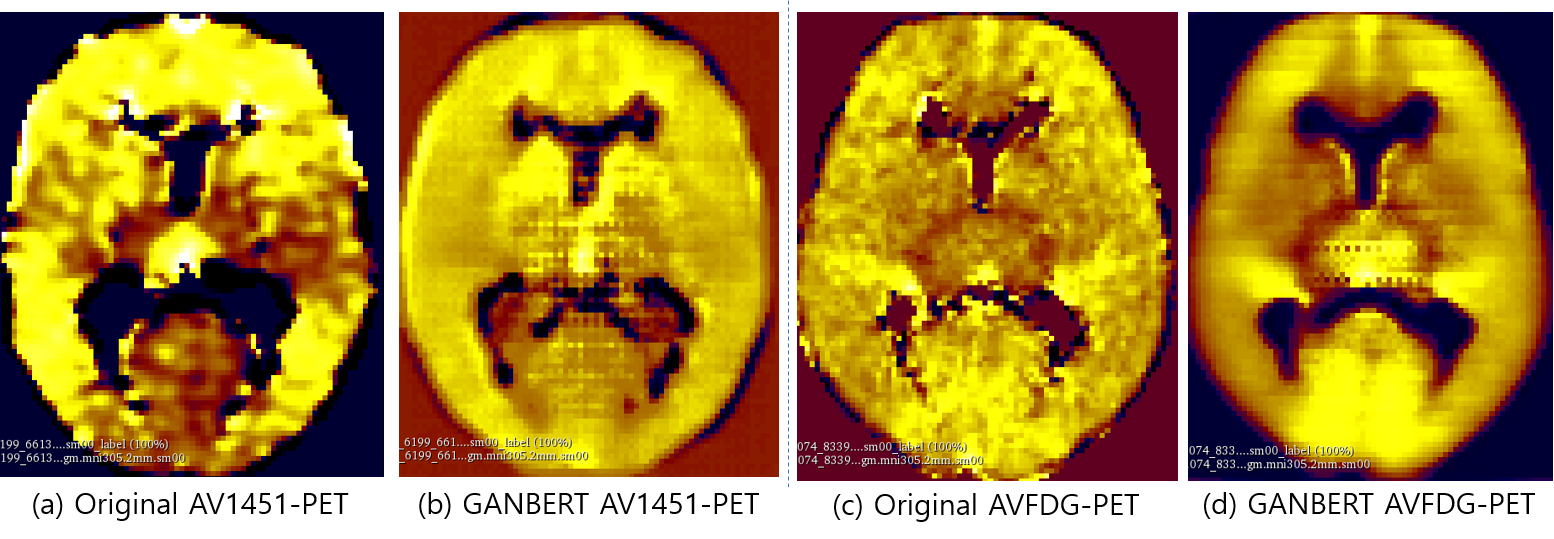}
\caption{Examples of AV1451 and FDG (a, c) PET and GANBERT synthesized (b, d).}
\label{fig:av1451_fdg_pet_gen}
\end{figure}

\subsection{Visual Examples of Generated PET Images}

Examples of AV45-PET images generated by GANBERT and \texttt{pix2pix} are compared with the original PET image in Figure~\ref{table:results_av45_pet_measure}.
GANBERT manages to generate close to the original AV45-PET image while \texttt{pix2pix} fails to generate a reasonable image.
Some more examples of AV1451- and FDG- PET images generated by GANBERT are compared with the original PET images in Figure~\ref{table:results_av1451_fdg_pet_measure}.
Overall, images to close to the original PET images are generated by GANBERT in all three PET imaging techniques tested.
The image quality of the center of the brain region is not as good as the other region, and it is suspected due to \textit{(1)} differences in MRI and PET images, and \textit{(2)} lack of training data.

\section{Conclusion}

We demonstrate synthesizing Positron Emission Tomography (PET) in original Standardized Uptake Value Ratio (SUVR) values from the T1-weighted MRI image input.
We achieve this with minimal pre-processing with no manual adjustments, mapping 0 to 255 range MRI intensity values to \texttt{-100}\,-\,\texttt{1000} in floating point values.


%
U-Net-like generator is combined and trained with BERT with NSP and MLM objectives, where NSP acts as GAN discriminator and MLM encouraging generator to reproduce highly biased intensity values.



\section*{Acknowledgement}
The authors would like to thank Seonjoo Lee of Columbia University Medical Center for the discussion and help in data pre-processing.


\bibliographystyle{splncs04}
\bibliography{ganbert}

\clearpage

\appendix

\section{Appendices}
\label{sec:appendix}

\subsection{Tanh and Tanhshrink}

\begin{figure}[h]
\centering
\includegraphics[width=.6\linewidth]{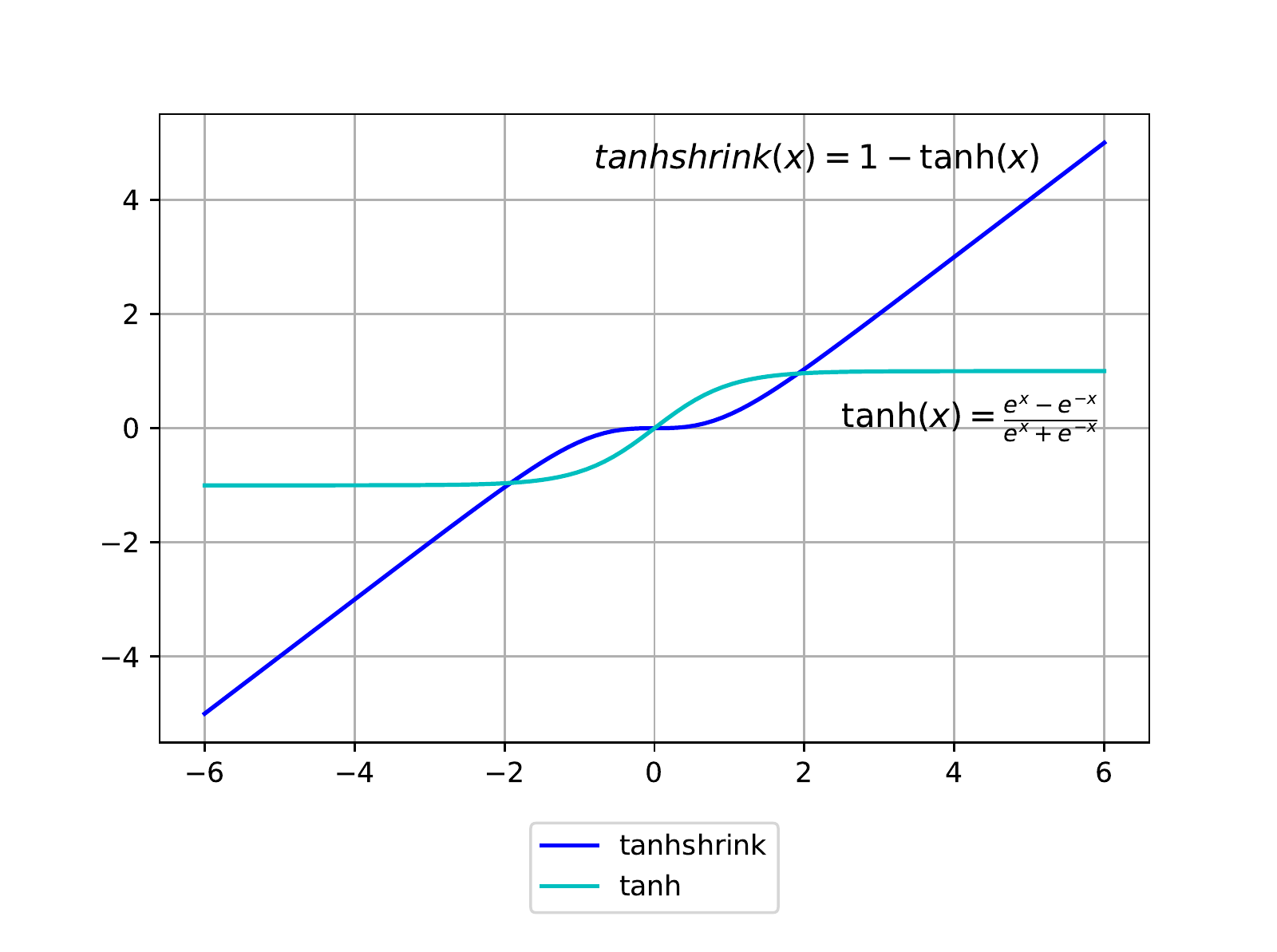}
\caption{\texttt{tanh} and \texttt{tanhshrink}.}
\label{fig:tanhshrink}
\end{figure}

The behaviors of \texttt{Tanh} and \texttt{Tanhshrink} activation functions are shown in Figure~\ref{fig:tanhshrink}.
While \texttt{Tanh} saturates -1 or 1, \texttt{Tanhshrink} does not.
We use \texttt{Tanhshrink} to reproduce the highly-biased intensity value range of PET SUVR.

\begin{figure}
\centering
\includegraphics[width=1\linewidth]{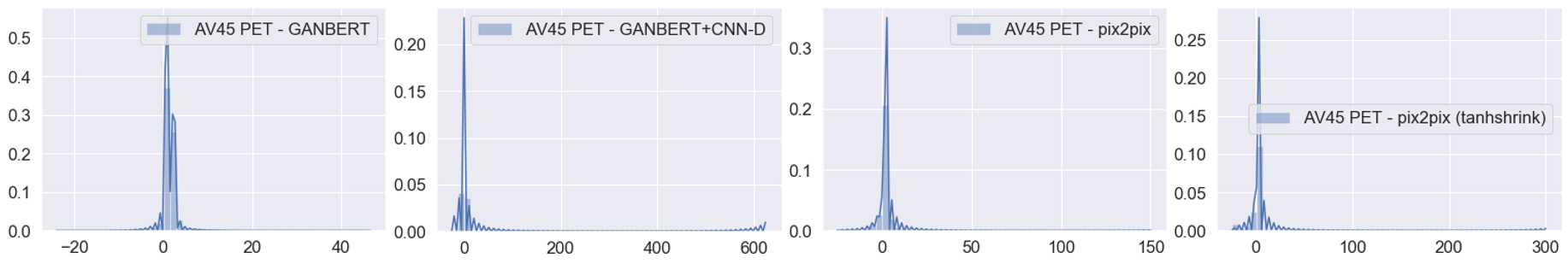}
\caption{Distribution of the image intensity values in PET images generated by \textit{(1)} GANBERT; \textit{(2)} GANBERT with additional CNN-Discrimator; \textit{(3)} \texttt{pix2pix} with \texttt{tanh} activation; \textit{(4)} \texttt{pix2pix} with \texttt{tanhshrink} activation.}
\label{fig:data_distribution_gen}
\end{figure}

\subsection{Restoring the Original PET Intensity Range}

Distribution of the generated PET images are shown in Figure~\ref{fig:data_distribution_gen}.
With GANBERT the original PET SUVR value range is closely preserved from negative to $\sim$50 range.
PET SUVR images floating-point values that are highly centered around zero, and it is important to preserve the details.
Yet, these small values are easily lost during deep learning training, especially when usual \texttt{tanh} or \texttt{sigmoid} activation are used.
Even when using \texttt{tanhshrink} activation, without GANBERT training they are hardly restored.
The distribution of PET generated by GANBERT with additional CNN-D shows wide intensity value range, but the details around zero are less well preserved.

\subsection{Training Details}


We perform distributed training on eight NVIDIA DGX-1 with 8$\times$ V100 GPUs for about 500 epochs, taking about a day.
Due to the GPU memory limit training batch size is 2, and we adopt gradient checkpointing~\cite{chen2016training} to reduce memory footprint.
That is still small, so we apply \textit{gradient accumulation}, accumulating gradients over two batches, making the training to be effectively with batch size of 4 for each GPU.
We use Adam~\cite{kingma2014adam} optimizer with learning ragte $1e-4$ and linear warm-up proportion using 5\% of the training data.

\end{document}